\documentclass[twocolumn]{article}
\usepackage[a4paper, total={180mm, 243mm}]{geometry}
\usepackage[utf8]{inputenc}
\usepackage{amssymb}
\usepackage{lipsum}
\usepackage{graphicx}
\usepackage{siunitx}
\usepackage[version=4]{mhchem} 
\usepackage{titling}
\usepackage{circledsteps}   
\usepackage{subcaption}
\usepackage{soul}
\usepackage[switch]{lineno} 
\modulolinenumbers[5]

\DeclareCaptionLabelSeparator{custom}{\textbar}
\DeclareCaptionFormat{custom}{\textsf{\textbf{#1 #2} \small #3}}
\newcommand\authormark[1]{\textsuperscript{#1}}

\makeatletter
\newcommand{\showfontsize}{\f@size{} pt}
\makeatother

\usepackage[
    backend=biber,
    style=nature,
    url=false,
    isbn=false,
    eprint=true,
    date=year,
    maxnames=99
]{biblatex}
\newcommand{\citet}[1]{\cite{#1}}
\newcommand{\citep}[1]{\cite{#1}}

\DeclareRobustCommand{\ion}[2]{%
  \relax
  \ifmmode
    \ifx\testbx\f@series
      {\mathbf{#1\,\mathsc{#2}}}
    \else
      {\mathrm{#1\,\mathsc{#2}}}
    \fi
  \else
    \textup{#1\,{\mdseries\textsc{#2}}}%
  \fi
 }

\definecolor{Myblue}{cmyk}{0.8,0.6,0,0.1}

\title{Mid-infrared continua via spectral broadening and difference frequency generation in a nanophotonic lithium niobate waveguide}

\author{Markus Ludwig\authormark{1,2,*},
        Furkan Ayhan\authormark{3},
        Thibault Voumard\authormark{1,4},
        Weichen Fan\authormark{1}, \\
        Mahmoud A. Gaafar\authormark{1,5,6},  
        Victor Brasch\authormark{7} ,
        Luis G. Villanueva\authormark{3},
        Tobias Herr\authormark{1,8,**}
}

\date{%
    \small $^1$Deutsches Elektronen-Synchrotron DESY, Notkestr. 85, 22607 Hamburg, Germany \\
    \small $^2$Presently: University of Luxembourg, 162a Avenue de la Faïencerie, L-1511 Luxembourg, Luxembourg, and Institute for Advanced Studies, University of Luxembourg, Campus Belval, L-4365 Esch-sur-Alzette, Luxembourg \\
    \small $^3$École Polytechnique Fédérale de Lausanne (EPFL), 1015 Lausanne, Switzerland \\
    \small $^4$Presently: Centre Suisse d’Électronique et de Microtechnique CSEM, Rue Jaquet-Droz 1, 2002 Neuchâtel, Switzerland \\
    \small $^5$Presently: Technology Innovation Institute TII, Abu Dhabi, United Arab Emirates\\
    \small $^6$Presently: Department of Physics, Faculty of Science, Menoufia University, Egypt\\
    \small $^7$Q.ANT GmbH, Handwerkstraße 29, 70565 Stuttgart, Germany\\
    \small $^8$Physics Department, Universität Hamburg UHH, Luruper Chaussee 149, 22607 Hamburg, Germany\\
    \small $^*$markus.ludwig@uni.lu \quad
    \small $^{**}$tobias.herr@desy.de
}

\begin{document}

\maketitle


\textbf{
Periodically poled thin film lithium niobate waveguides provide simultaneous access to efficient second and third order nonlinear processes, enabling broadband generation of coherent laser light. Here, we demonstrate the generation of a broadband mid-infrared continuum in a nanophotonic lithium niobate waveguide pumped by a telecom-wavelength femtosecond source. Specifically, our dual-stage design includes both third-order nonlinear spectral broadening followed by a dedicated periodically poled waveguide section performing efficient broadband intrapulse difference frequency generation. Driven by sub-100~fs pulses with $\sim$200~pJ pulse energy, the generated mid-infrared light covers wavelengths from 3200 to 4800\,nm. Cascaded harmonic generation also extends the spectrum into the visible and ultraviolet domains, resulting in an overall spectral bandwidth ranging from 350 to 4800\,nm. 
}

\begin{figure*}[t]%
\centering
\includegraphics[width=140mm]{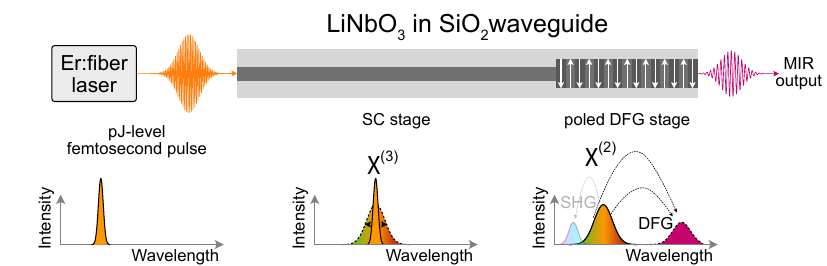}
\caption{\textbf{Concept of MIR generation in a dual stage lithium niobate waveguide.} A femtosecond pulse from a telecom-wavelength fiber laser experiences spectral broadening via supercontinuum generation (SC) due to the third-order nonlinearity in the first stage. In the second stage, mid-infrared light is generated via difference frequency generation (DFG), facilitated by a quasi-phase matched second-order nonlinearity; in addition second harmonic generation (SHG) and higher harmonics generation may occur.}\label{fig1}
\end{figure*}

Access to coherent broadband light sources and frequency combs in the mid-infrared (MIR) \cite{sorokina_solid-state_2003, schliesser_mid-infrared_2012, zhang_advances_2024} enables fundamental studies and catalyzes several technological applications. Among these are the exploration of ultrafast phenomena \cite{lanin_time-domain_2014,pires_ultrashort_2015,schoenfeld_nonlinear_2024}, molecular fingerprinting and functional group sensing, environmental monitoring, trace gas sensing, and precision frequency metrology \cite{diddams_optical_2020, picque_frequency_2019, fortier_20_2019}. Relying on established femtosecond laser sources in the near-infrared (NIR), access to broadband mid-infrared (MIR) has predominantly been achieved in several ways: supercontinuum generation based on third order nonlinear effects \cite{sylvestre2021RecentAdvancesSupercontinuum, zorin_advances_2022, bres2023SupercontinuumIntegratedPhotonics}, as well as effects based on the second order optical nonlinearity: optical parametric oscillation (OPO) \cite{muraviev_massively_2018}, optical parametric generation (OPG) \cite{petrov_parametric_2012,hu_highly_2023}, and difference frequency generation (DFG) \cite{bonvalet_generation_1995,kaindl_femtosecond_nodate,erny_mid-infrared_nodate,gaida_watt-scale_2018,ycas_high-coherence_2018, lind_mid-infrared_2020,ritzkowsky_passively_2023,bournet_maximizing_2024}. 

MIR combs generated via DFG, especially intrapulse DFG (IDFG), can be of zero carrier-envelope offset frequency \cite{baltuska_controlling_2002, krauss_all-passive_2011}, enabling the precise control of optical waveforms in the time domain and an elegant approach to absolute (SI time standard referenced) spectroscopy.

MIR light generation via IDFG requires the generation of a broadband spectrum that covers a large frequency interval $\mathrm{\Delta\nu}$. IDFG can then in principle generate frequencies ranging from zero up to $\mathrm{\Delta\nu}$, where in practice material absorption often limits the achievable MIR bandwidth. To produce MIR light with wavelengths of $\mathrm{3\,\mu m}$ or longer, a pump pulse with a bandwidth of at least $\mathrm{100\,THz}$ is required. 
Previous work has already utilized bulk $\mathrm{LiNbO_3}$ \cite{lesko_six-octave_2021, hoghooghi_broadband_2022} and large mode-area waveguides~\cite{kowligy_mid-infrared_2018,kowligy_mid-infrared_2020} as a nonlinear medium for IDFG into the MIR. 
Its large second order nonlinear coefficient and its ferroelectricity, enable efficient quasi-phase matched second order nonlinear processes and light conversion across multiple octaves, particularly when implemented as nanophotonic waveguides with sub-$\mu m^2$ cross-section \cite{wang_ultrahigh-efficiency_2018,wu_visible--ultraviolet_2024, ludwig2024UltravioletAstronomicalSpectrograph}. Previous implementations of IDFG in $\mathrm{LiNbO_3}$ have relied on spectral broadening in a nonlinear fiber \cite{iwakuni_generation_2016,lesko_six-octave_2021, hoghooghi_broadband_2022,kowligy_mid-infrared_2018,hettel_compact_2024} or nanophotonic waveguides \cite{kowligy_mid-infrared_2020}, where the resulting supercontinuum is launched either into a bulk periodically poled $\mathrm{LiNbO_3}$ (PPLN) crystal or large area PPLN waveguide. This approach decouples the nonlinear broadening from the IDFG process, allowing the preparation of short and broadband input pulses before entering the dedicated DFG stage. In aluminum nitride waveguides, which like lithium niobate have large second and third order nonlinearities but without the possibility of periodic poling, spectral broadening and IDFG have been observed in the same waveguide, indicating potential for compact MIR sources \cite{hickstein_ultrabroadband_2017}.

Here, we explore the integration of both spectral broadening (supercontinuum generation) and IDFG into a single monolithic two-stage nanophotonic $\mathrm{LiNbO_3}$ waveguide (see Figure \ref{fig1}). The IDFG section in particular incorporates broadband quasi phase matching via periodic poling. 
Based on this structure we demonstrate MIR light generation from an erbium-doped femtosecond fiber laser: Tailoring the group velocity dispersion (GVD) optimizes the efficiency and spectral extent of the supercontinuum generation, whereas group index matching and quasi-phase matching are key for maximizing the IDFG efficiency. 

\begin{figure*}[t]%
\centering
\includegraphics[width=\textwidth]{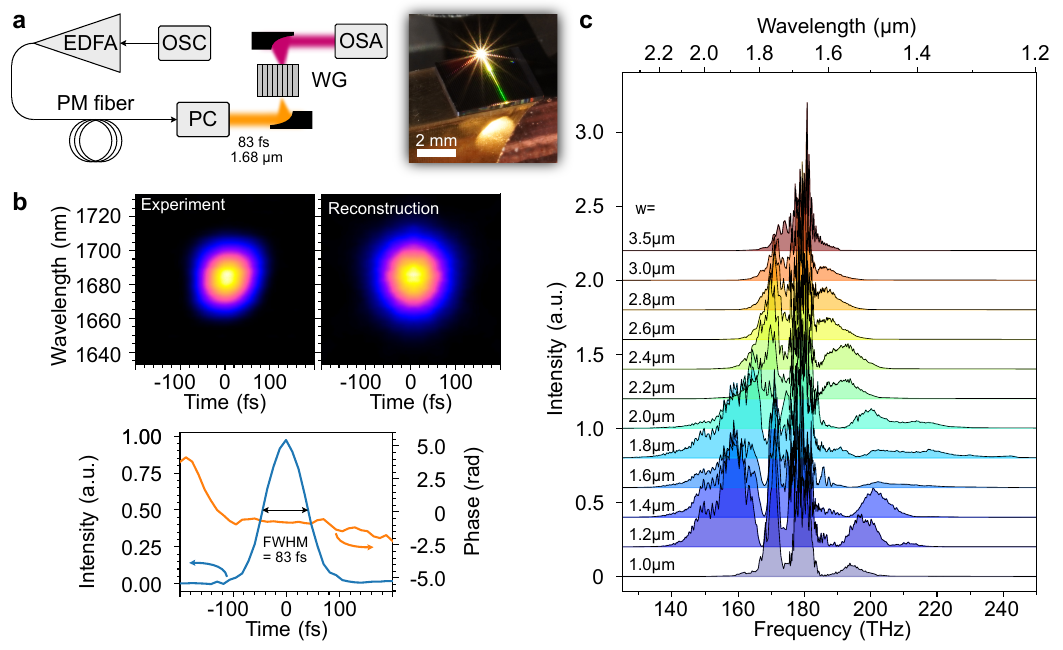}
\caption{\textbf{Supercontinuum generation in an unpoled LiNbO$_\mathbf{3}$ waveguide.} \textbf{a)} Experimental setup. A telecom-wavelength femtosecond oscillator (OSC) seeds an Er:fiber amplifier (EDFA). The pulses propagate through 1\,m of standard polarization maintaining (PM) fiber. A prism compressor (PC) isolates the resulting soliton pulses. Parabolic mirrors couple light into and out of the LiNbO$_{3}$ waveguides. Optical spectrum analyzers (OSA) are used to measure the output spectra. The photograph shows a LiNbO$_{3}$ waveguide in operation. \textbf{b)} Experimental FROG trace (left) and reconstructed trace (right) indicate a bandwidth limited pulse of 83\,fs at a central wavelength of 1680\,nm. \textbf{c)} Supercontinuum generation in LiNbO$_{3}$ waveguides without poling at an on-chip pulse energy of 300\,pJ. Generated supercontinua as a function of waveguide (top) width. A bandwidth exceeding 100\,THz at the 10\% intensity level can be achieved.
}\label{fig2}
\end{figure*}

Figure \ref{fig2}\,a) illustrates the experimental test setup. As a pump source we use a homemade Er:fiber laser \cite{kafka_mode-locked_1989,brida_ultrabroadband_2014} with a repetition rate of $\mathrm{40\,MHz}$. After amplification in a core-doped Er:fiber amplifier up to $\mathrm{300\,mW}$, we obtain pulses with a maximum energy of $\mathrm{2.3\,nJ}$. Propagation in 1\,m of standard polarization maintaining (PM) 1550\,nm fiber results in the formation of 83\,fs Raman-shifted soliton pulses with a central wavelength of $\mathrm{1680\,nm}$. Spectral filtering 
via a free-space silicon prism compressor configured for zero net dispersion, provides well-defined light pulses; this setups also permits tuning of the optical pulse energy via a variable neutral density filter without affecting pulse duration or chirp. Off-axis parabolic mirrors couple the pulses into the waveguide and collimate the output light achromatically. Assuming a coupling efficiency per facet of $\mathrm{13\,\%}$ (see below), this provides a maximum of 300\,pJ on-chip pulse energy. Figure \ref{fig2}\,b), shows the reconstructed pulse parameters based on second harmonic generation frequency resolved optical gating (SHG-FROG) \cite{delong_frequency-resolved_1994}.

\begin{figure*}[h!t]%
\centering
\includegraphics[width=\textwidth]{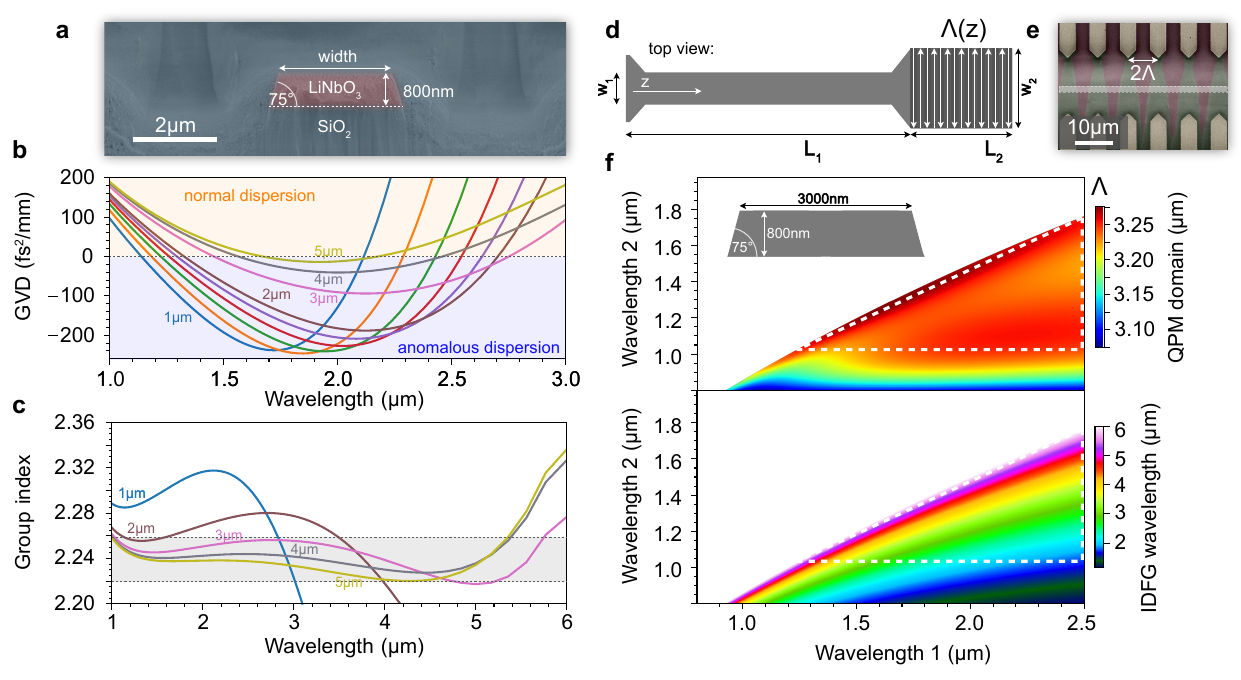}
\caption{\textbf{Waveguide design} \textbf{a)} False color scanning electron micrograph of a waveguide cross section. \textbf{b)} Calculated group velocity dispersion as a function of wavelength for different waveguide (top) widths. \textbf{c)} Calculated group index as a function of wavelength for different waveguide widths. \textbf{d)} Layout of the two-stage waveguide. In a narrow section with length L$_1=4$\,mm a supercontinuum is generated. Subsequent IDFG is performed in a wider, periodically poled section of length L$_2=1$\,mm for minimal temporal walk-off and improved long-wavelength mode confinement. \textbf{e)} False color scanning electron micrograph indicating the poled material before etching. Shown are the electrodes spaced by a poling period $2\Lambda$ and the resulting domain inversion after poling. The horizontal light gray bar indicates the location of the later etched waveguide. \textbf{f)} Calculated quasi-phase matching for a waveguide width of 3000\,nm. Top: Color-coded domain width (half poling period) as a function of two input wavelengths (1\&2) driving IDFG. Bottom: Resulting output IDFG wavelengths as a function of the two input wavelengths. The nearly constant color code inside the dashed triangle (top graph) indicates a nearly uniform requirement for the poling period for IDFG output wavelengths between $\mathrm{3-6\,\mu m}$ (bottom graph).
}\label{fig3}
\end{figure*}

First, to design the initial spectral broadening stage, we fabricate a series of $\mathrm{SiO_2}$ cladded, unpoled $\mathrm{LiNbO_3}$ on insulator (LNOI) waveguides of different (top) widths, a fixed side wall angle of $75^\circ$, a film thickness of $\mathrm{800\,nm}$ (see SEM image in Figure \ref{fig3} a)) and overall length of $\mathrm{5\,mm}$~\cite{ayhan2025FabricationPeriodicallyPoled}. 
The crystal axis is oriented such that it aligns with the transverse electric (TE) mode. With these waveguides, we perform supercontinuum generation with the highest available pulse energy of 300\,pJ to identify the optimum between dispersion-limited spectral broadening and enhanced nonlinearity due to the mode confinement. Figure~\ref{fig2}\,c) shows the resulting spectra on a linear intensity scale as a function of frequency. The widest spectral broadening is achieved for waveguide widths between $\mathrm{1-2\,\mu m}$, reaching spectral bandwidths beyond 100\,THz. In our case, we select a waveguide width of $\mathrm{1.2\,\mu m}$, providing a broadband spectrum while still avoiding spectral overlap between the expected IDFG comb and the broadened fundamental comb. In addition, we simulate waveguide dispersion via the finite element method (FEM) based on commonly available bulk material data \cite{zelmon_infrared_1997,malitson_irving_h_interspecimen_nodate}. Figure \ref{fig3} b) illustrates the resulting GVD for different waveguide widths. Anomalous dispersion - a requirement for broadband supercontinuum generation via self-compression - can be achieved for all widths. Values between $\mathrm{1-2\,\mu m}$ maximize the spectral width of the anomalous dispersion region centered close to the pump wavelength, which agrees with the experimental findings.

Second, for optimal IDFG efficiency, temporal overlap of all frequency components along the DFG stage is important. Figure \ref{fig3} c) shows the group index for different waveguide geometries based on the dispersion simulation. The lowest variation in the group index of less than 1\% across a spectral bandwidth of $\mathrm{1-5\,\mu m}$ is achieved for waveguide widths between $\mathrm{3-5\,\mu m}$, and, to maximize nonlinearity, we choose a width of $\mathrm{3\,\mu m}$. 
In addition to group index matching, phase matching is a critical requirement, which we approach here through periodic poling: Figure \ref{fig3} f) visualizes the quasi-phase matching (QPM) conditions for the poling domain width $\Lambda = \frac{\pi}{\Delta k}$ as a function of possible combinations of two input wavelengths. The upper panel indicates the domain width (half poling period) while the lower panel indicates the resulting DFG wavelength. Interestingly, the QPM domain width for DFG output wavelengths $\mathrm{3-6\,\mu m}$ (see dashed triangle in Figure \ref{fig3} g) ) can be achieved with a nearly uniform domain width of $\mathrm{ 3.25\pm0.05\,\mu m}$ (poling period of $\mathrm{ 6.5\pm0.1\,\mu m}$). At the cost of a possibly reduced conversion efficiency, we choose a linearly chirped poling, covering periods from $\mathrm{ 6.3-6.8\,\mu m}$, to accommodate potential fabrication imperfections. We also note that, even though unintended, the QPM period for SHG phase matching is relatively similar at $\mathrm{ 6.2\,\mu m}$. The final design of the two-stage waveguide is illustrated in Figure \ref{fig3}\,d). Out of the total waveguide length of 5\,mm, we dedicate the first 4\,mm to spectral broadening, and 1\,mm to IDFG.
The waveguide width at the chip's input facets is $\mathrm{2.5\, \mu m}$ and tapers over a distance of $\mathrm{200\, \mu m}$ down to $\mathrm{w_1=1.2\,\mu m}$. The output facet width equals the width of the DFG stage $\mathrm{w_2=3\,\mu m}$ and the taper between the narrow and wide waveguide sections extends over $\mathrm{250\, \mu m}$. 

\begin{figure*}[h!t]%
\centering
\includegraphics[width=\textwidth]{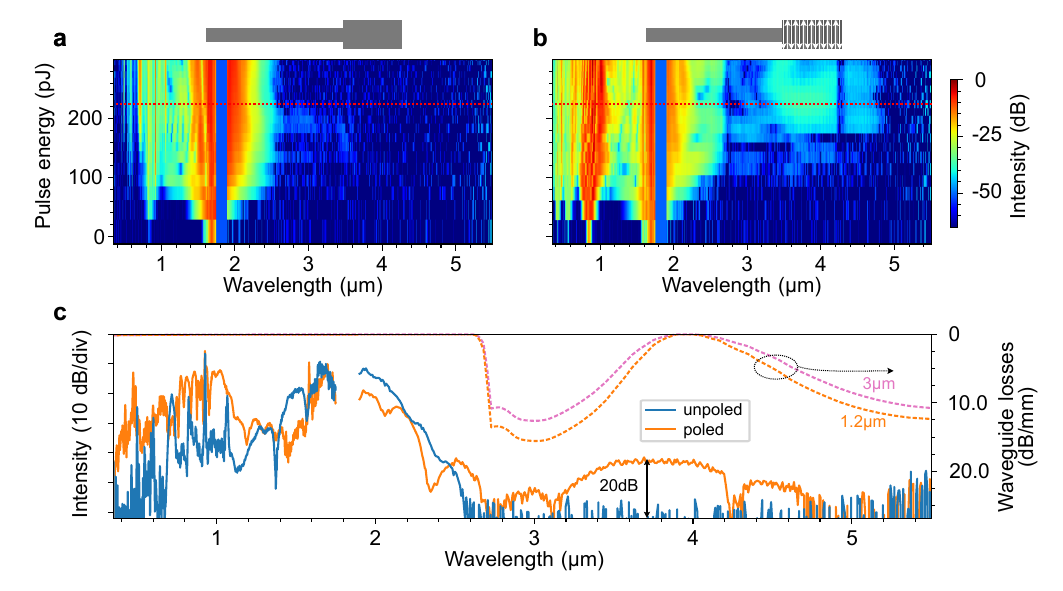}
\caption{\textbf{MIR comb generation via IDFG} \textbf{a)} Output spectra of an unpoled waveguide as a function of estimated on-chip input pulse energy. \textbf{b)} Output spectra of an identical waveguide geometry as in a) but with periodic poling in the IDFG stage. Spectral slices at 236\,pJ indicated by the red dotted lines are shown in panel \textbf{c)}. In contrast to the unpoled waveguide, where no light generation beyond 2.5\,$\mu$m wavelength is observed, the poled waveguide shows significant MIR light generation, 20dB above the noise floor. The dashed lines indicate waveguide losses calculated for the two involved waveguide widths.
}\label{fig4}
\end{figure*}

Next, we pump the waveguide with our femtosecond source and record the output spectrum as a function of input pulse energy using two optical spectrum analyzers (Yokogawa AQ6377 and AQ6374). The resulting spectra are shown in Figures \ref{fig4}a) and b) and are recorded with identical waveguide geometry under the same coupling conditions, with the sole distinction that the sample in Figure \ref{fig4}a) is unpoled while the one in Figure \ref{fig4}b) is poled. The unpoled waveguide does not generate detectable light beyond a wavelength of $\mathrm{2.5\,\mu m}$. In contrast, the poled waveguide exhibits significant light conversion into a wide window between $\mathrm{3-4.8\,\mu m}$. The roll-off beyond $\mathrm{4.8\,\mu m}$ wavelength correlates well with the absorption expected from the PECVD-grown silica cladding, as shown in Figure \ref{fig4}c) (right axis). The absorption is modeled via FEM for both the the SC section ($\mathrm{1.2\,\mu m}$ width) and the DFG section ($\mathrm{3\,\mu m}$ width), and found to exceed 10 dB/mm at wavelength longer than $\mathrm{5\,\mu m}$. The OH absorption band between $\mathrm{2.5-3.5\,\mu m}$ results in a  transmission loss of 12~dB along the DFG section. A relatively narrow absorption feature in the supercontinua at $\mathrm{4.2\,\mu m}$ can be attributed to $\mathrm{CO_2}$ absorption. While unintended in the design, the spectra also witness a highly efficient conversion into the second, third and fourth harmonics. We attribute this to higher poling order and mode order phase matching and non-symmetric poling (i.e. not strictly 50:50 duty cycle due to fabrication imperfections). Figure \ref{fig4}\,c) shows spectral slices corresponding to the dotted red lines of Figures \ref{fig4}\,a),b) at an on chip pulse energy of $\mathrm{236\,pJ}$. The MIR portion in the poled sample is enhanced by at least 20~dB compared to the unpoled structure, consistent with IDFG. 
Based on a broadband thermal powermeter, the total collimated off-chip output power, corresponding to the spectrum in Figure \ref{fig4}\,c) for the poled waveguide, amounts to $\mathrm{1.22\,mW}$. Comparison with the off-chip input power measured on the same thermal powermeter allows us to determine the transmission to be at least $\mathrm{1.7\%}$. Assuming a symmetric coupling, this translates to a coupling efficiency of $\mathrm{13\%}$ per facet. 
The IDFG-based MIR spectrum, spectrally separated from the pump supercontinuum, could in principle constitute a frequency comb with zero carrier-envelope offset frequency. However, recent studies show that spectral overlap between the fundamental supercontinuum and its second harmonic (which is the case here) can generate interleaved combs with non-zero offset frequency across the entire spectrum~\cite{fan2025SpectralDynamicsBroadband}; thus further studies are needed to clarify whether the spectral structure of the MIR spectra.

In conclusion, we demonstrate MIR generation from a telecom-wavelength erbium fiber laser through supercontinuum generation and subsequent IDFG, both realized in a single 5 mm nanophotonic $\mathrm{LiNbO_3}$ waveguide. Quasi-phase matching via periodic poling enables MIR emission spanning $\mathrm{3200\,nm}$–$\mathrm{4800\,nm}$ at input pulse energies of about 200 pJ. Future work may investigate the spectral structure of the IDFG output, in particular its carrier-envelope offset frequency. Overall, our results establish a new link between the telecom and MIR bands, opening opportunities for integrated MIR photonics and future research.


\subsection*{Acknowledgements}
The lithium niobate waveguides were fabricated in the EPFL Center of MicroNanoTechnology (CMi). The work was supported through the Maxwell computational resources operated at DESY. We thank Yokogawa Deutschland GmbH for lending us a mid-infrared optical spectrum analyzer (AQ6377).

\subsection*{Funding}
\small
This project has received funding from the Swiss National Science Foundation (Sinergia BLUVES CRSII5\_193689), the European Research Council (ERC, No 853564), the European Innovation Council (EIC, 101046920), through the Helmholtz Young Investigators Group VH-NG-1404 and PIER, the partnership of Universität Hamburg and DESY (grant ID PIF-2022-09); the work was supported through the Maxwell computational resources operated at DESY.


\subsection*{Data availability}
\small
The datasets analyzed during the current study are available from the corresponding authors upon reasonable request.


\subsection*{Competing interests}
\small
We declare that none of the authors have competing interests.

\printbibliography

\end{document}